\documentclass[%
 reprint,
 amsmath,amssymb,
 aps,
 prd,
]{revtex4-2}

\usepackage{hyperref}
\usepackage{graphicx}
\usepackage{dcolumn}
\usepackage{bm}
\usepackage{color}
\usepackage{aas_macros}

\begin{document}

\title[SN seismology]{Unveiling the Nature of Gravitational-Wave Emission in Core-collapse Supernovae with Perturbative Analysis}

\author{Shuai Zha}\email{zhashuai@ynao.ac.cn}
\affiliation{Yunnan Observatories, Chinese Academy of
Sciences (CAS), Kunming 650216, China}
\affiliation{Key Laboratory for the
Structure and Evolution of Celestial Objects, CAS, Kunming 650216,
China}
\affiliation{International Centre of Supernovae, Yunnan Key Laboratory,
Kunming 650216, China}
\author{Oliver Eggenberger Andersen}
\author{Evan P. O'Connor}
\affiliation{The Oskar Klein Centre, Department of Astronomy, Stockholm University, AlbaNova, SE-106 91 Stockholm, Sweden}%

\date{\today}

\begin{abstract}
Gravitational waves (GWs) can provide crucial information about the central engines of core-collapse supernovae (CCSNe). In order to unveil the nature of GW emission in CCSNe, we apply perturbative analyses with the same underlying equations as simulations to diagnose oscillations of the proto-neutron star (PNS) during $\sim$1~s postbounce. In the pseudo-Newtonian case, we find that radial profiles of GW emission match well between the perturbative analysis with $l=2$ and simulations inside the PNS at \emph{any} frequency and time. This confirms that the GW emission of CCSNe arises from the global PNS oscillations in the perturbative regime. Based on this, we solve for the discrete eigenmodes with a free PNS surface and tentatively identify a set of $g$ modes and the $f$ mode contributing to the peak GW emission. We also offer a possible explanation for the power gap in the GW spectrum found in simulations that lies at the frequency with vanishing cumulative emission of the PNS. Our results enhance the predictive power of perturbative analyses in the GW signals of CCSNe.
\end{abstract}

\maketitle

\section{Introduction} \label{sec:intro}
Core-collapse supernovae (CCSNe), the explosive death of massive stars with initial masses greater than $\sim8$\,$M_\odot$, are potential astrophysical sources of gravitational waves (GWs) with frequencies from hundreds to thousands of Hz (see \citep{abdikamalov22,mezzacappa24} for recent reviews). Current \citep{ligo,virgo,kagra} and next-generation \citep{NEMO,ET,CE} ground-based GW detectors may detect CCSNe taking place in the Milky Way and nearby galaxies \citep{2016PhRvD..93d2002G,2021PhRvD.104j2002S,bailes21,2022Galax..10...70S}. Unlike electromagnetic signals, GWs carry direct information of the CCSN central engine without intervention by the stellar envelope, so its detection will boost our understanding of the CCSN physics, e.g., the core structure of progenitors \citep{2014PhRvD..90d4001A,warren20,
2021ApJ...914...80P} and the explosion mechanism(s) \citep{2009CQGra..26t4015O,2021ApJ...914...80P,2023PhRvD.107j3015V,powell23}, as well as the nuclear physics inside CCSN cores, in particular the high-density finite-temperature equation of state (EOS) \citep{marek09,2021ApJ...923..201E,sotani21,2023PhRvL.131s1201J,2017PhRvD..95f3019R,2020PhRvL.125e1102Z}. 

In general, the detection of GWs requires some preknowledge of the waveform emitted from a source \citep{speri22}. The study of GW emission from CCSNe has evolved from semi-analytical calculations of collapsing ellipsoids \citep{shapiro78} to 2-dimensional (2D) hydrodynamical simulations of collapsing rotating polytropic stars \citep{moenchmeyer91,zweger97} before the year 2000. With the advance of high-performance computing in the new millennium, nowadays one can simulate CCSNe starting from realistic progenitor models  while using finite-temperature and composition dependent EOSs and capturing neutrino transport using sophisticated solvers, all while in full 3D (e.g., \citep{andresen17,oconnor18,mezzacappa20}). 3D magnetohydrodynamic CCSN models are also available now to provide GW waveforms for magnetorotational supernovae \citep{moesta14,martin21,powell23a,shibagaki23}, potentially associated with long gamma-ray bursts.   These theoretical investigations are forming an invaluable knowledge base for the CCSN GW detection.  

In CCSN models with non-rotating progenitors, GW emission usually arises after the core bounce due to the stiffening of EOS. The postbounce GWs have a typical waveform after $\sim100$\,ms with stochastically varying amplitudes, while the peak GW frequency exhibits a characteristic ramp-up trend from 100's to 1000's of Hz with time. \textit{Murphy et al.} \citep{murphy09} and \textit{M\"uller et al.} \citep{muller13} showed that the GW peak frequency is in close resemblance to the buoyancy frequency near the proto-neutron star (PNS) surface in Newtonian and relativistic simulations, respectively. Using perturbative analysis, several groups \citep{torres18,torres19,morozova18,sotani20} identified that the peak GW emission corresponds to $l=2$ gravity ($g$) and fundamental ($f$) mode oscillations of the PNS by matching its frequency with those of perturbative eigenmodes.  Such perturbation analyses complement the computationally intensive multi-D CCSN simulations and provides a more intuitive way to understand the complex numerical results. Furthermore, with 25 1D simulations and a perturbative analysis, \textit{Torres-Forn{\'e} et al.} \citep{torres19b} derived universal relations between the frequencies of a few important eigenmodes and the PNS surface gravity as well as the mean density inside the shock. \textit{Sotani et al.} \citep{sotani21} later provided an alternative formula with dependence solely on the PNS mean density. If these universal relations are reliable, they will assist the search of CCSN GWs concealed within the detector noises and eventually the inference of important CCSN parameters \citep{bizouard21,bruel23}.

However, discrepancy remains among these studies \citep{torres18,torres19,morozova18,sotani20,wolfe23} in the exact association between perturbative eigenmodes ($g_1$, $g_2$, $f$ and etc.) and the peak GW emission in CCSN simulations, in part due to different approximations (Cowling vs. non-Cowling) and the choice of boundary conditions. This hinders the predictive power of perturbative analyses. \textit{Westernacher-Schneider} \citep{ryan2020} pointed out that the mode functions match well between the perturbative analysis and simulation that share the same underlying hydrodynamic equations. They demonstrated this approach using simulation data recorded with a high cadence during $t\in[30,50]$\,ms postbounce when the typical GW emission has not emerged. 

In this work, we follow the methodology of Ref.~\cite{ryan2020} to analyze the GW emission in the later epochs, when the shock has expanded to $\sim100-200$\,km and neutrino-driven convection develops in between the PNS and shock. We show that the radial profiles of GW emission inside the PNS match well with those from a consistent perturbative analysis at \emph{any} frequency and time, \emph{not} only for the eigenmodes. This suggests that the typical CCSN GW emission mainly comes from the $l=2$ oscillations of the perturbed PNS. Based on this fact, we obtain the eigenmodes using the usual boundary condition, i.e. vanishing Lagrangian pressure perturbation at the PNS surface. We tentatively associate the peak GW emission with a set of $g$ modes and the $f$ mode below and above a frequency gap as identified in Refs.~\cite{morozova18,2021ApJ...923..201E}. We further notice that the gap may be related to a zero point of total GW emission inside the PNS with a small variation in its frequency during the $\sim1$\,s postbounce. This encourages further attention on this gap to understand its origin as well as its relation to the PNS properties.

Our paper is organized as follows. We describe the adopted CCSN models for the GW signals and simulation data in \S\,\ref{sec:model}. In \S\,\ref{ssec:perturb} we present the method for the perturbative analysis. \S\,\ref{ssec:match} shows the matching between the radial profiles of GW emission from simulation and perturbative analysis. \S\,\ref{ssec:eigen} presents the association of the peak GW frequency with the perturbative eigenmodes. We discuss the possible origin of the GW emission gap in \S\,\ref{ssec:gap}. We conclude our findings in \S\,\ref{sec:conclu}.

\section{Core-collapse supernova models} \label{sec:model}
We use the results of 2D axisymmetric CCSN simulations performed in Ref.~\cite{2021ApJ...923..201E} for our analysis. They are simulated with Newtonian hydrodynamics in \texttt{FLASH (v.4)} \citep{FLASH} implemented with a multi-group and multi-species two-moment neutrino transport scheme and an approximate relativistic gravitational potential \citep{oconnor18a}. We focus on the 5 non-rotating CCSN models using a 20-$M_{\odot}$ solar-metallicity progenitor model \citep{woosley07} and a set of Skyrme-type EOSs \citep{sroeos18,schneider19}. 3 models vary the effective nucleon mass (models m0.55, m0.75, m0.95), and 2 models fix the effective nucleon mass as m0.75 but vary the isoscalar incompressibility modulus (models m0.75\_k200 and m0.75\_k260). We refer the readers interested in the details about the numerical setups and EOSs to the original paper \citep{2021ApJ...923..201E}.

\begin{figure}
    \centering
    \includegraphics[width=0.49\textwidth]{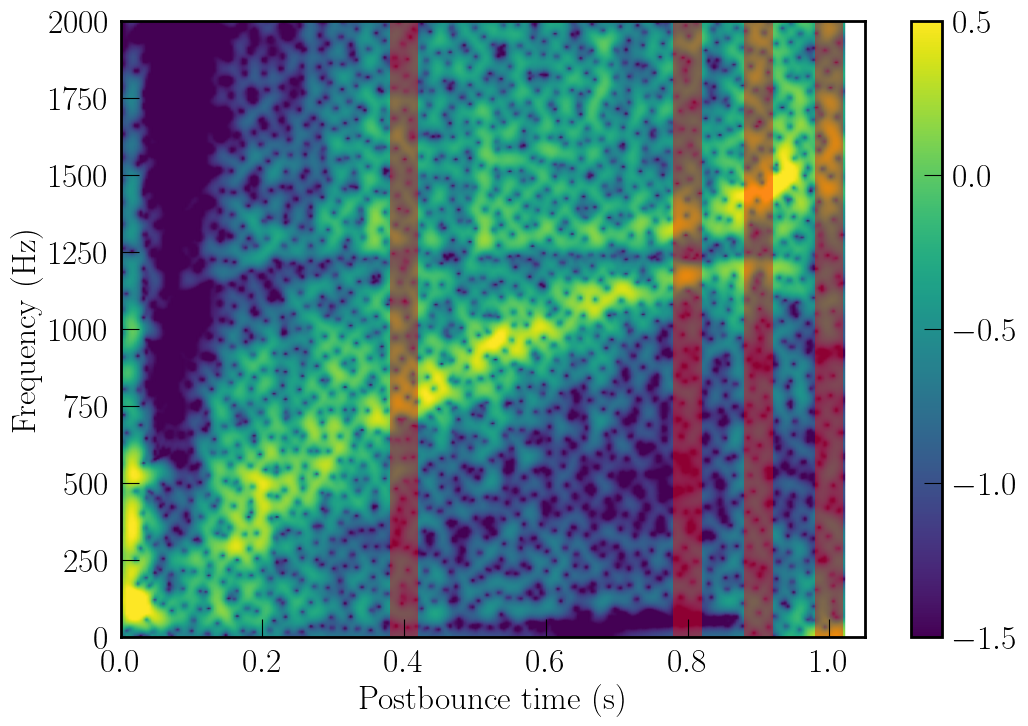}
    \caption{Gravitational-wave spectrogram in the logarithmic scale for the model m0.75. The power spectral density is normalized to the maximum value after 0.1\,s postbounce. The red shaded regions indicate the re-simulated interval with a high cadence of data recording ($2\times10^{-5}$\,s). These regions are centered at 0.4, 0.8, 0.9 and 1.0\,s postbounce with an interval of 40\,ms.   }
    \label{fig:gw_spec}
\end{figure}

The dimensionless GW strain is extracted from simulations with the quadrupole formula, \citep{finn90}
\begin{equation} \label{eq:hplus}
    h_+(t) = \frac{3}{2}\frac{G}{Dc^4} \frac{d^2}{dt^2} I_{zz},
\end{equation}
where we assume an observer located at the equator and a distance $D$ away from the source. $I_{zz}$ is the only independent component of the quadrupole moment in the axisymmetric case
\begin{equation} \label{eq:Izz}
    I_{zz} = \int dV (z^2 - \frac{1}{3}r^2) \rho.
\end{equation}
where $dV$ is the volume element, $r$ is the spherical radius, and $\rho$ is the source density (rest-mass in our Newtonian case).  The integration in Eq.~\ref{eq:Izz} is over the whole star to get the overall GW strain. Fig.~\ref{fig:gw_spec} shows the postbounce GW spectrogram derived from $h_+(t)$ in the model m0.75 by the short-time Fourier transform. One can clearly see the general trend of increasing peak GW frequency as mentioned in \S\,\ref{sec:intro}. A power gap is also visible nearby $\sim1250$\,Hz with slowly varying frequency throughout the simulated postbounce time.

Snapshots recorded with a high cadence are necessary for a direct comparison of the spatial contribution to the GW emission between simulations and a perturbative analysis \citep{ryan2020}. Therefore, we re-simulate the model m0.75 to get 1 snapshot every $2\times10^{-5}$\,s in four 40-ms intervals centered at 0.4, 0.8, 0.9, and 1.0\,s postbounce, as indicated by the red shaded regions in Fig.~\ref{fig:gw_spec}. Then we get the radial profile of GW strain by,
\begin{equation} \label{eq:hspatial}
    \begin{aligned}
        & h_+(t,r)  = \frac{3}{2}\frac{G}{Dc^4} \frac{d^2}{dt^2} I_{zz}(r), \\
        & I_{zz}(r) = \int_{r-\Delta r/2}^{r+\Delta r/2} dV (z^2 - \frac{1}{3}r^2) \rho,
    \end{aligned}
\end{equation}
where the integration is over a spherical shell with a thickness of $\Delta r=1\,{\rm km}$ to get $I_{zz}(r)$. We warn that if partial integration is applied to avoid the numerical differentiation in time for $h_+(t,r)$ \citep{finn90}, one needs to take care of the surface terms (see the appendix in Ref.~\cite{2021ApJ...923..201E}). 

\begin{figure}
    \centering
    \includegraphics[width=0.49\textwidth]{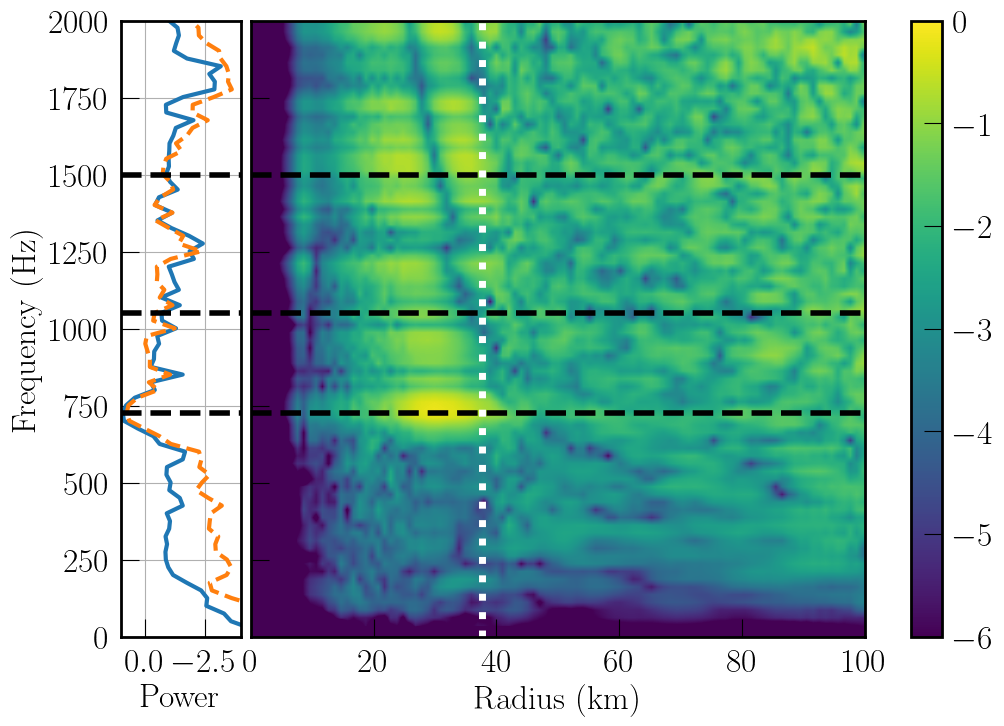}
    \caption{The 2D colormap plots the radial profile of gravitational-wave power spectral density in the logarithmic scale for the model m0.75 in the 40-ms interval centered at 0.4\,s postbounce. The vertical dotted line denotes the proto-neutron star (PNS) surface with $\rho=10^{11}\,{\rm g\,cm}^{-3}$ and the horizontal black dashed lines denote the frequencies (725, 1050 and 1500\,Hz) chosen for matching the perturbative functions. The lines plotted in the left panel show the overall gravitational-wave spectrum for the regions inside 100\,km (solid) and the PNS (dashed).  }
    \label{fig:rf_power}
\end{figure}

We focus our analysis on the spatial contribution to GW emission in frequency space, i.e. $\tilde{h}_+(f,r)$ derived by the fast Fourier transform from $h_+(t,r)$ of the 40-ms intervals. The 2D colormap in Fig.~\ref{fig:rf_power} illustrates the normalized $|\tilde{h}_+(f,r)|$ of the re-simulated interval centered at 0.4\,s postbounce. The PNS surface is marked with the white vertical line, which is defined as where the spherically averaged density equals $10^{11}\,{\rm g\,cm^{-3}}$. Characteristic structures exist in $|\tilde{h}_+(f,r)|$ inside the PNS above $\sim 600\,$Hz and will be compared to the perturbative functions in \S\,\ref{ssec:match} at 3 selected frequencies marked by the horizontal dashed lines. In the left panel, we plot the overall GW spectrum for the regions inside 100\,km (solid) and the PNS (dashed). Above $\sim 600\,$Hz, the PNS region dominates the GW emission, especially around the peak GW frequency $\sim725$\,Hz. Note that sometimes the dashed line is above the solid line because the contribution of outer regions may increase or decrease the total power depending on the relative sign of $\tilde{h}_+(f)$ inside and outside the PNS.

\section{Perturbative analysis} 
\subsection{Methodology} \label{ssec:perturb}
To be consistent with the pseudo-Newtonian simulations, we follow the method of Ref.~\cite{ryan2020} to perform a perturbative analysis with our background CCSN data. Here we only list the essential equations for completeness and refer the interested readers  to Ref.~\cite{ryan2020} for a more thorough derivation. Note that we use geometric units $G=c=1$ throughout this section. 

In our adopted simulations, the gravitational potential $\Phi$ is calculated from the Poisson's equation,
\begin{equation} \label{eq:poisson}
    \nabla^2 \Phi = 4\pi\rho,
\end{equation}
by a multipole expansion method with spherical harmonic orders up to $l=16$ \citep{couch13}. The monopole term of $\Phi$ is then modified to approximate general-relativistic effects using the Case A formula of Ref.~\cite{2006A&A...445..273M}. As pointed out in Ref.~\cite{ryan2020}, this relativistic correction is irrelevant for the perturbative analysis of GW emission because the lowest order contribution arises from quadrupolar accelerations ($l=2$). Therefore, the purely Newtonian perturbative equations with $l=2$ \cite{christensen1991solar,ryan2020} are consistent with the kind of pseudo-Newtonian CCSN simulations considered here.

To linearize the hydrodynamic equations and the Poisson's equation, we use the Ansatz that the Eulerian perturbative value of each variable $u$ can be written as
\begin{equation} \label{eq:ansatz1}
    \delta u = \delta \hat{u}(r) Y_l e^{-i\sigma t}, 
\end{equation}
where $u$ stands for the density $\rho$, the pressure $P$, the gravitational potential $\Phi$ and the Eulerian radial displacement $\xi_r$. Thus $\delta \hat{u}$ stands for the frequency-domain perturbative variables, i.e. $\delta \hat{\rho}$, $\delta \hat{P}$, $\delta \hat{\Phi}$ and $\eta_r$. $Y_l$ is the spherical harmonic function, which for our axisymmetric case is the Legendre polynomial with order $l$. $\sigma=2\pi f$ is the angular frequency. The Eulerian tangential displacement $\xi_\theta$ is written differently, as
\begin{equation} \label{eq:ansatz2}
    \delta \xi_{\theta} = \frac{\eta_\bot(r)}{r^2} \partial_\theta Y_l e^{-i\sigma t}.
\end{equation}
Hereafter, we omit the $r$ dependence in $\delta \hat{u}(r)$ for conciseness. We set $l=2$ in our calculations due to the dominant role of the quadrupolar perturbation in GW emission.

\begin{figure}
    \centering
    \includegraphics[width=0.48\textwidth]{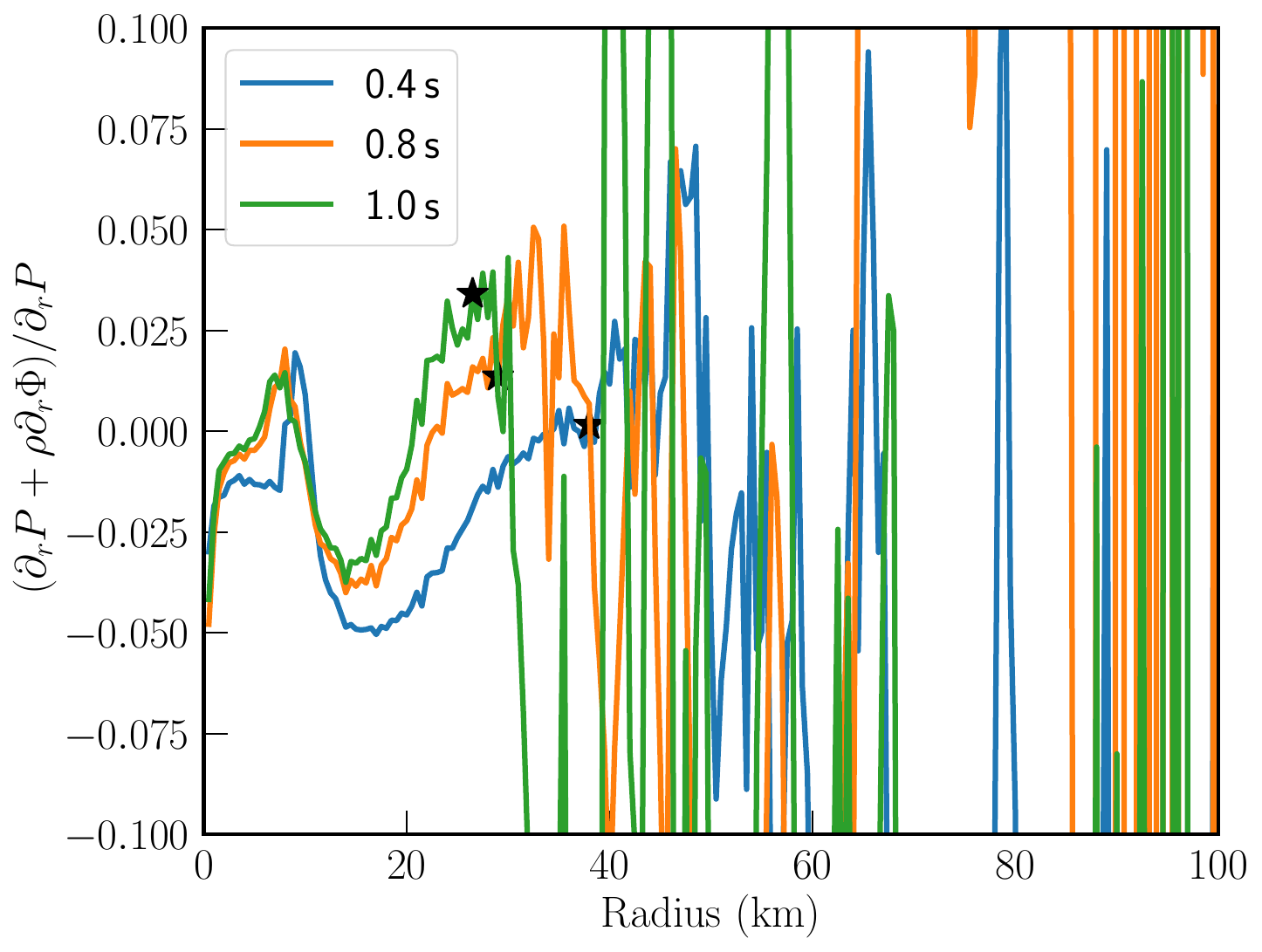}
    \caption{Deviation from the hydrostatic condition $\partial_r P=-\rho \partial_r \Phi$ at 3 postbounce times, 0.4\,s, 0.8\,s and 1.0\,s in the model m0.75. The black stars mark the proto-neutron star surface defined as where the spherically averaged density equals $10^{11}\,{\rm g\,cm^{-3}}$.}
    \label{fig:static}
\end{figure}

With the adiabatic condition $\Delta P/\Delta \rho=c_s^2$ ($\Delta$ stands for Lagrangian perturbation) and substituting Eqs.~\ref{eq:ansatz1} and \ref{eq:ansatz2} into the hydrodynamic and Poisson's equations, we can get the perturbative equations as
\begin{equation} \label{eq:perturb}
    \partial_r \vec{w} = \mathbf{A} \vec{w},
\end{equation}
with
\begin{equation}
    \vec{w} = (\eta_r,\eta_\bot,\delta \hat{\Phi}, F)^T,
\end{equation}
and
\begin{widetext}
    \begin{equation}
    \mathbf{A} =\begin{pmatrix}
       -2/r-\partial_r \ln P/\Gamma_1 &  -\sigma^2/c_s^2+l(l+1)/r^2 & 1/c_s^2 & 0\\
       1-N^2/\sigma^2 & -\mathcal{B} & \mathcal{B}/\sigma^2 & 0 \\
       0 & 0 & 0 & 1 \\
       -4\pi\rho\mathcal{B} & 4\pi\rho\sigma^2/c_s^2 & -4\pi\rho/c_s^2+l(l+1)/r^2 & -2/r \\
    \end{pmatrix}.
    \label{eq:matrix}
\end{equation}
\end{widetext}
Here, $F\equiv \partial_r \delta \hat{\Phi}$, $\Gamma_1$ is the adiabatic index, and $c_s^2=\Gamma_1 P/\rho$ is the speed of sound in the Newtonian limit. $\mathcal{B}\equiv \partial_r \ln\rho-(1/\Gamma_1)\partial_r \ln P$ is the Schwarzschild discriminant, $N^2=\tilde{G}\mathcal{B}$ is the Brunt-V\"ais\"al\"a frequency squared with $\tilde{G}\equiv\partial_r P/\rho=-\partial_r \Phi$. We note that the assumption of hydrostatics ($\partial_r P=-\rho \partial_r \Phi$, see Fig.~\ref{fig:static}) is valid for the PNS within $\sim5\%$ while it becomes much worse outside PNS, in the neutrino-heating convective region. Also, $v_r$ becomes markedly non-zero outside the PNS. These facts suggest that the perturbative equations become invalid outside the PNS. We emphasize again that Eq.~\ref{eq:perturb} is only consistent with our pseudo-Newtonian simulations for $l\neq0$.

We use a 4-stage Runge-Kutta scheme to integrate Eq.~\ref{eq:perturb} from a small non-zero radius $r_0=dr/5$, where $dr=0.5\,$km is the radial step for the numerical integration. We linearly interpolate the spherically averaged background profiles from simulations to get hydrodynamic variables at any given radius. We impose the following regularity conditions at $r_0$
\begin{equation}
    \begin{aligned}
        \eta_r &= A_0 r^{l-1},\quad\quad \eta_\bot=\frac{A_0}{l}r^l, \\
        \delta \hat{\Phi} &= C_0r^l,\quad\quad \partial_r \delta \hat{\Phi}=lC_0r^{l-1},
    \end{aligned}
\end{equation}
where we set $A_0=10^{-5}$ and solve for $C_0$ with the Newton-Raphson method to fulfill the outer boundary condition 
\begin{equation}
    \Big[\partial_r \delta \hat{\Phi}+\frac{l+1}{r}\delta \hat{\Phi}\Big]\Big|_{r=R} =0,
\end{equation}
where a large enough $R$ would not affect the perturbative solution \cite{ryan2020} and we set $r=100\,$km. We have checked that this is adequate for not affecting our conclusions.

Normally another outer boundary condition is applied to obtain a set of discrete \emph{eigenmodes}. One common choice is vanishing Lagrangian perturbation in pressure, i.e. $\Delta P=0$ at the PNS surface usually defined with a specific density cut \citep{morozova18,sotani20,torres18}. Refs.~\citep{torres19,ryan2020} claim a more reasonable boundary condition as vanishing Eulerian radial displacement ($\eta_r=0$) at the shock radius. The choice of this boundary condition affects the identification of the eigenmodes responsible for the GW emission and we will present the results of eigenmodes in \S\,\ref{ssec:eigen}.

\subsection{Matching perturbative solutions with simulations} \label{ssec:match}
In reality, one can solve the perturbative equation (Eq.~\ref{eq:perturb}) for a mode function at \emph{any} frequency without imposing the second boundary condition. We demonstrate with the model m0.75 that the spatial profiles of GW emission match well at \emph{any} frequency between the simulation (cf. Fig.~\ref{fig:rf_power}) and the perturbative analysis, particularly inside the PNS region.

The contribution of a mode to GW emission ($\tilde{h}_+(f,r)$) can be expressed as \citep{thorne69} 
\begin{equation} \label{eq:gw_perturb}
    \tilde{h}_+(f,r) \propto \int_{r-\Delta r/2}^{r+\Delta r/2} dV \delta \hat{Q}(f,r) ,
\end{equation}
where $\delta \hat{Q}=r^2\delta \hat{\rho}$ is the perturbed quadrupole moment. $\delta \hat{\rho}$ can be obtained from solutions of the perturbative equation (Eq.~\ref{eq:perturb}) by
\begin{equation}
    \delta \hat{\rho} = \rho \Big( \frac{\sigma^2}{c_s^2}\eta_\bot - \frac{\delta \hat{\Phi}}{c_s^2} - \mathcal{B} \eta_r\Big).
\end{equation}

We illustrate the comparison between the simulation and perturbative analysis for the time interval centered at 0.4\,s in Fig.~\ref{fig:match} and at 0.8\,s, 0.9\,s and 1.0\,s in Appendix~\ref{app:match}. We select 3 frequencies for the comparison with one in the vicinity of the peak GW emission and the other two randomly picked. The frequencies from left to right in Fig.~\ref{fig:match} are denoted as horizontal dashed lines in Fig.~\ref{fig:rf_power} from bottom to top, respectively. 

\begin{figure*}
    \centering
    \includegraphics[width=0.95\textwidth]{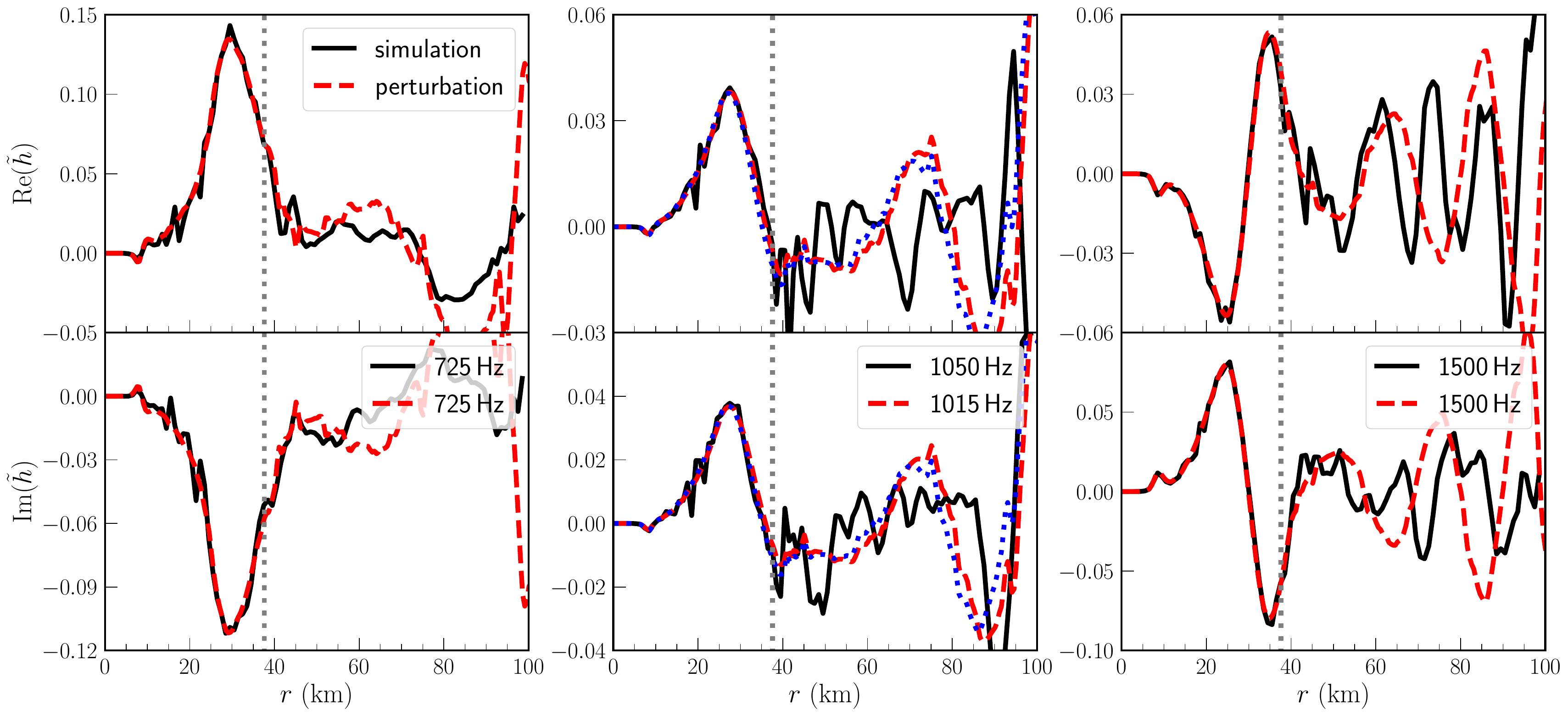}
    \caption{Examples of matching radial profiles of gravitational-wave power spectral density between the simulation (black solid lines, also see Fig.~\ref{fig:rf_power}) and perturbative analysis (red dashed lines) for the model m0.75 in the $0.04$-s interval centered at 0.4\,s postbounce. The frequencies in the simulation are 725, 1050 and 1500\,Hz from left to right. Note that we have searched for the best-fit perturbative function in a $\pm100$\,Hz window centered at the simulated frequency. The blue dotted lines show the perturbative function with the same frequency as the simulation when the best-fit frequency differs from that. Note that we multiply the perturbative function with a frequency dependent constant to match the simulated signals.}
    \label{fig:match}
\end{figure*}

As the Fourier component $\tilde{h}_+(f,r)$ in the simulation is transformed from the GW waveform of a 40-ms interval, its frequency resolution is at best $\sim$25\,Hz. Instead of fixing at the same frequency, we opt to search for the best-fit perturbative mode in nearby frequencies. Therefore, we solve for the perturbative modes in the $\pm100$\,Hz window centered at the simulated frequency with a step of 5\,Hz. Fig.~\ref{fig:match} shows that within the PNS surface (indicated by gray dotted lines), both real and imaginary parts of $\tilde{h}_+(f,r)$ from the simulation matches almost perfectly with the perturbative $\delta Q(f,r)$ (Eq.~\ref{eq:gw_perturb}) multiplied by a constant. Together with other intervals presented in Appendix~\ref{app:match}, the discrepancy of frequency between the simulation and perturbative analysis is always within 50\,Hz and can be better than 25\,Hz when $\tilde{h}_+(f,r)$ has a large amplitude with respect to numerical noise. However, mismatch appears outside the PNS and becomes more prominent farther away.

The above results indicate that the GW emission of CCSNe predicted by simulations mostly comes from the PNS oscillations in the perturbative regime.  This supports the use of a consistent perturbative analysis for the interpretation of the GW signals predicted from sophisticated nonlinear simulations. It also implies that the perturbative equation is only valid within the PNS so that a proper outer boundary condition should be chosen not too far away from the PNS surface.

\subsection{Eigenmodes relevant for the peak gravitational-wave emission} \label{ssec:eigen}
The ultimate goal of performing perturbative analysis is to identify the true perturbative \emph{eigenmodes} associated with the peak GW emission. Based on this, pragmatic universal relations can be built between the peak GW frequency and important physical parameters \citep{torres19b,sotani21}. Below we examine the usual frequency matching routine to dig out the desired eigenmodes.

We obtain the eigenmodes with the commonly-used outer boundary condition, i.e. a free surface \citep{morozova18,sotani20}
\begin{equation} \label{eq:bc}
        \Delta P =\rho(\sigma^2\eta_\bot-\delta \hat{\Phi})+\eta_r \partial_r P=0.
\end{equation}
Our default choice is applying this condition at the tentative PNS surface where the spherically averaged density equals $10^{11}\,{\rm g\,cm^{-3}}$. We only search for the eigenmodes in between 250\,Hz and 2000\,Hz which is most relevant to the GW emission in our CCSN models. 

\begin{figure*}
    \centering
    \includegraphics[width=0.45\textwidth]{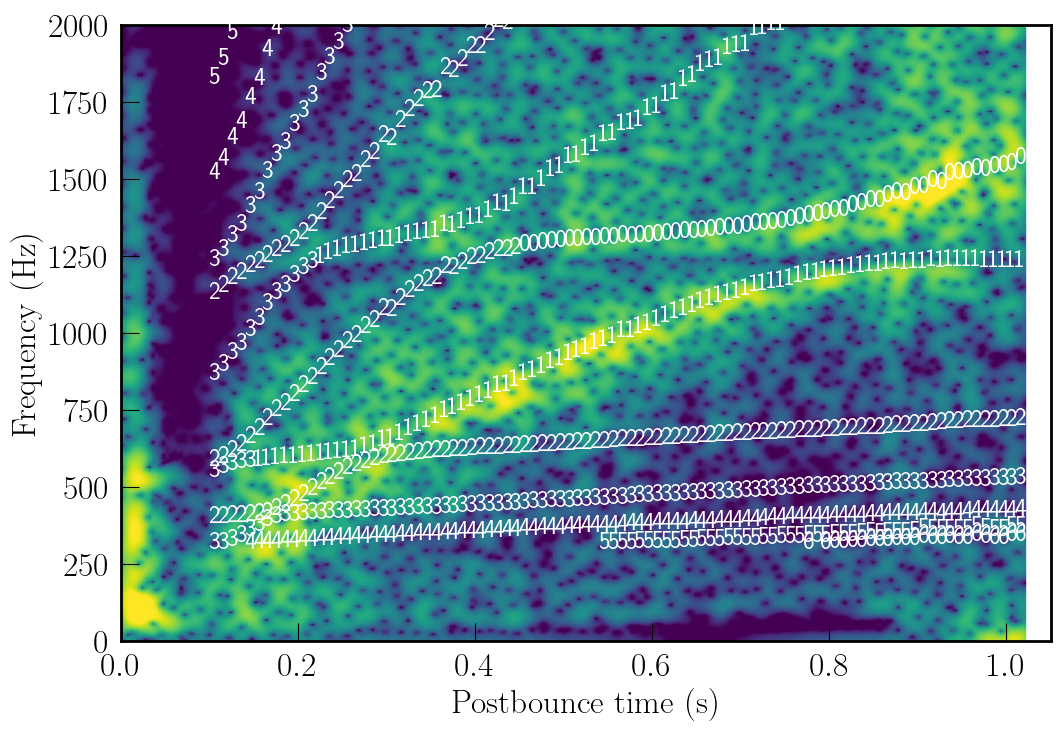}
    \hspace{0.025\textwidth}
    \includegraphics[width=0.45\textwidth]{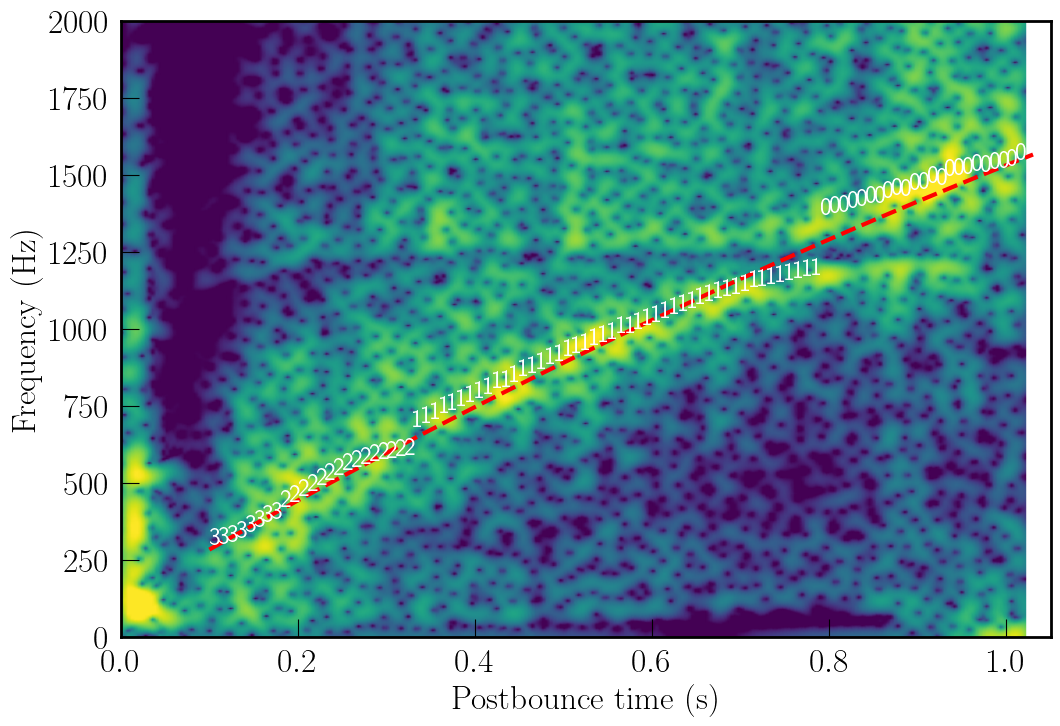}
    \caption{\textit{Left panel: } Gravitational-wave (GW) spectrogram seen in Fig.~\ref{fig:gw_spec} with the frequencies of perturbative eigenmodes over-plotted as digits denoting the corresponding numbers of radial nodes. \textit{Right panel: } Same as the left panel, but with the red dashed line denoting the quadratic fitting curve of the GW peak frequency as a function of postbounce time. The digits denotes the frequencies of perturbative eigenmodes closest to the red line.}
    \label{fig:spec+mode}
\end{figure*}

In Fig.~\ref{fig:spec+mode}, we plot frequencies of the resulting eigenmodes on top of the GW spectrogram for the model m0.75. The digits denote the numbers of radial nodes ($n$) in $\eta_r$ of the corresponding eigenmodes. Ref.~\citep{torres19} discussed about difficulties in assigning the nature of a mode as gravity ($g$), pressure ($p$) or hybrid ($h$). Here we follow the simplest Cowling method \citep{cowling41} to classify the eigenmodes. From the upper end, the eigenmodes with decreasing $n$ for a lower frequency are classified as $p_n$ modes, while from the lower end those with decreasing $n$ for a higher frequency are classified as $g_n$ modes. The eigenmodes in between $p$ and $g$ modes are classified as $h$ modes as Refs.~\citep{torres18,torres19}, and the eigenmode with $n=0$ is designated as the $f$ mode. The $f$ mode emerges at $\sim0.45$\,s postbounce and, after this time, well separates the $p$ and $g$ modes with frequencies above and below it. There are several crossings between eigenmodes with $n$ differing by 1.

In the right panel of Fig.~\ref{fig:spec+mode} we identify the eigenmodes that are responsible for the peak GW emission by the frequency matching method. The red dashed line shows a quadratic fitting curve of the peak GW frequency as a function of time after 0.1\,s postbounce \citep{2021ApJ...923..201E}. At every 10\,ms, we plot the eigenmode whose frequency is closest to this curve. Tentatively, the peak GW emission is associated with a set of $g_n$ modes while $n$ decreases consecutively by 1 at the crossing of modes. After $\sim0.8\,$s postbounce, the $f$ mode is responsible for the peak GW emission whose frequency is above the power gap near $\sim1250$\,Hz.

In Appendix~\ref{app:eigen}, we present similar results for other 2 models, m0.55 and m0.95. Here our frequency matching is based on a consistent perturbative analysis with the pseudo-Newtonian simulation, and the credibility comes from the excellent mode function matching within the PNS, cf. \S\,\ref{ssec:match}. Yet uncertainty remains in the choice of the outer boundary condition for solving the eigenvalue problem. As Refs.~\citep{morozova18,sotani19}, we have checked that a lower density cut of $10^{10}\,{\rm g\,cm^{-3}}$ for the PNS surface does not alter the eigenmode association, especially for the later times. Ref.~\citep{torres19} claimed that a more physical condition is $\eta_r=0$ at the shock radius. However, this condition is not pragmatic as in multi-D simulations the shock becomes highly aspherical at a radius of 100's of km. A conclusive association of eigenmodes with the peak GW emission awaits for further creative and rigorous investigations.

\subsection{The power gap}
\label{ssec:gap}
Another intriguing feature is the ``power gap" in the GW spectrum of CCSNe in the interval of $[1000,1300]$\,Hz as first noted by \textit{Morozova et al.} \citep{morozova18} in 2D axisymmetric simulations with \texttt{FORNAX}. This is also found in 3D simulations with \texttt{FLASH} \citep{oconnor18} and with \texttt{FORNAX} \citep{vartanyan23} but not in Refs.~\citep{mezzacappa20,mezzacappa23} with \texttt{CHIMERA}. Refs~\citep{morozova18,vartanyan23} attribute this power gap to the avoided crossing of $g$ and $f$ modes and to a trapped $g$ mode in the inner PNS core (within a radius of $\sim10$\,km). Here we offer an alternative explanation for the power gap with our perturbative analysis.

\begin{figure*}
    \centering
    \includegraphics[width=0.45\textwidth]{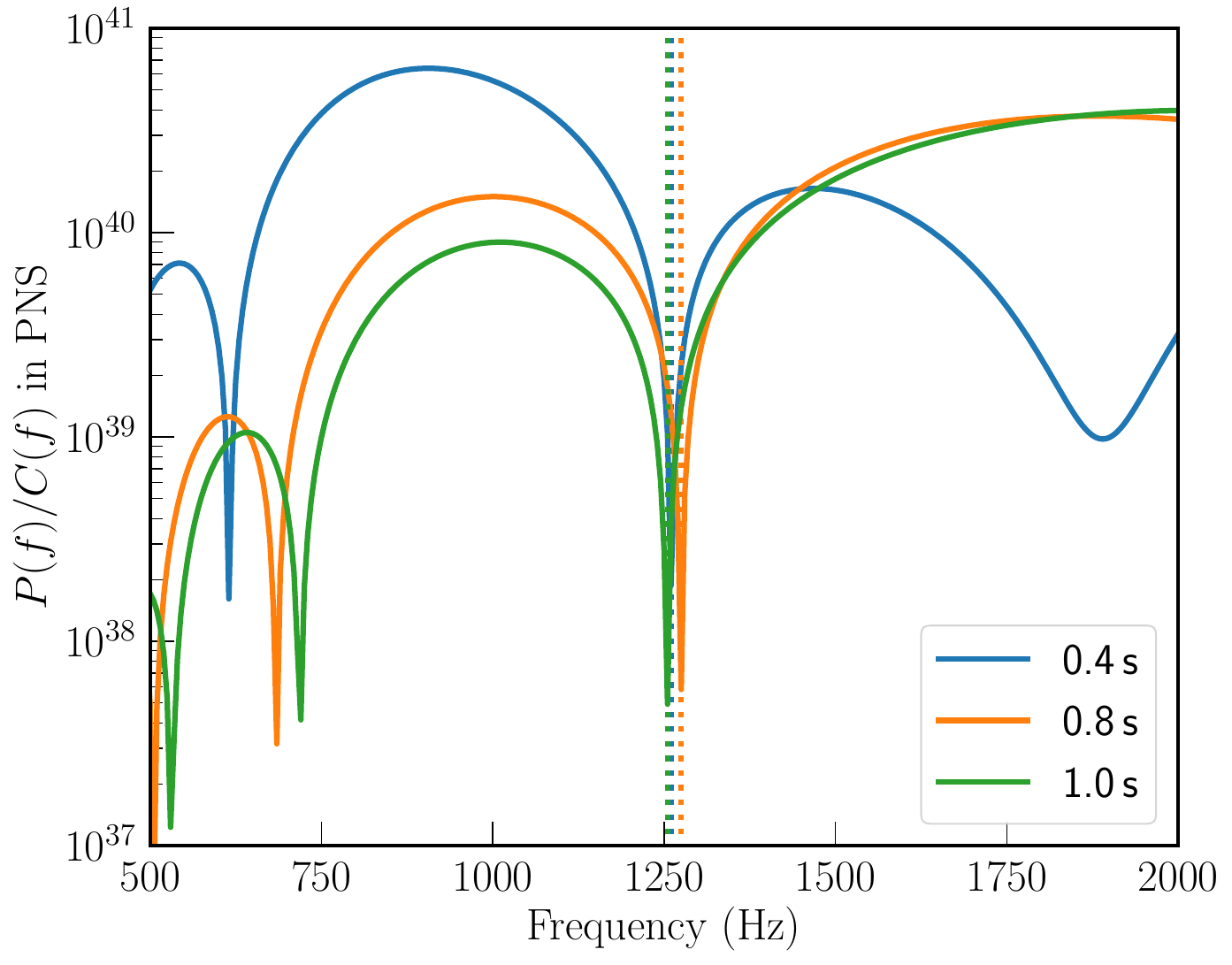}
    \hspace{0.025\textwidth}
    \includegraphics[width=0.45\textwidth]{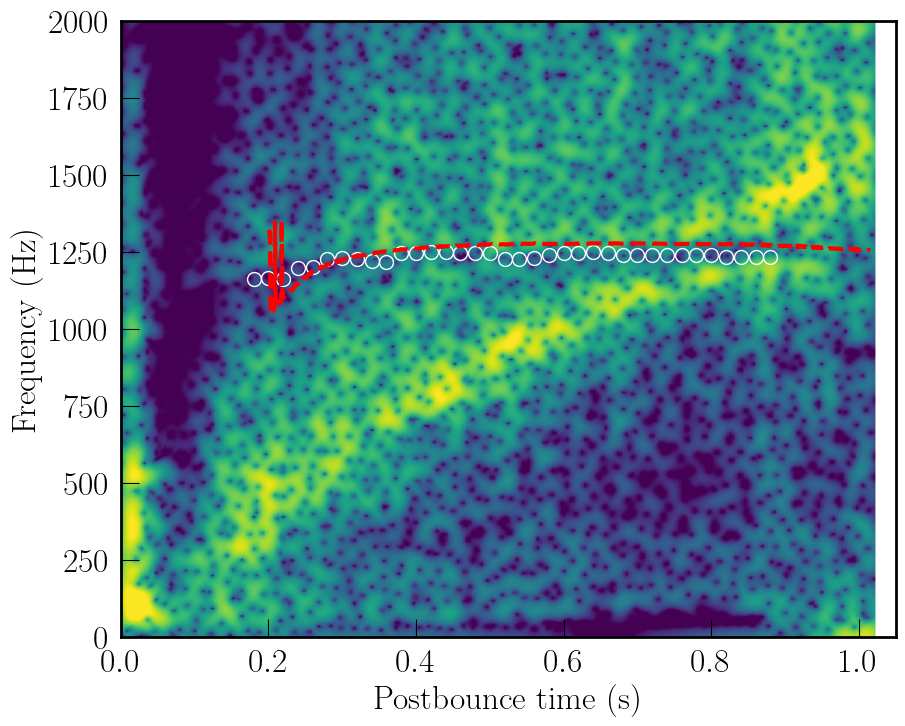}
    \caption{\textit{Left panel: } Total power ($P(f)/C(f)$ in Eq.~\ref{eq:htot}) of the perturbative mode functions inside the proto-neutron star (PNS) as a function of frequency at 0.4, 0.8 and 1.0\,s postbounce. The PNS surface is defined as the locus of $10^{11}\,{\rm g\,cm}^{-3}$. The vertical dotted lines denote the local minimum around the gap frequency in the simulation ($\sim1250 $\,Hz). \textit{Right panel: } Same as Fig.~\ref{fig:gw_spec}, but with the gap frequencies over-plotted as white open circles. The red dashed line denote the frequency for the local minimum power inside PNS as the vertical dotted lines in the left panel.}
    \label{fig:gap75}
\end{figure*}

As shown in \S\,\ref{ssec:match}, the GW emission profiles in the simulation well match with the perturbative functions at any frequency inside the PNS. Based on the fact that the PNS region generally dominates the GW emission (cf. the left panel of Fig.~\ref{fig:rf_power}), we compute the total power of GW emission inside the PNS from the perturbative functions by
\begin{equation} \label{eq:htot}
    P(f) = {C}(f) \int_{0}^{R_{\rm PNS}} dV \delta \hat{\rho} r^2, 
\end{equation}
where ${C}(f)$ is a frequency-dependent factor which relates to the ability of excitation at $f$ and cannot be determined by the perturbative analysis. In the left panel of Fig.~\ref{fig:gap75}, we plot the absolute value of this power (without the unknown ${C}(f)$) as a function of frequency at 0.4, 0.8 and 1.0\,s in the model m0.75. A local minimum is found near the gap frequency $f_{\rm gap}\sim1250$\,Hz in the power spectrum as marked by the vertical dotted lines. This local minimum results from a zero point in the total power as its sign changes. In the right panel of Fig.~\ref{fig:gap75} we plot the frequency of this local minimum in power ($f_{\rm min}$) as a function of time with the red dashed line on top of the corresponding GW spectrogram. The white open circles show $f_{\rm gap}$ determined from the local minimum within $[1000, 1300]$\,Hz in the GW spectrum derived from the waveform with a window of 200\,ms. After $\sim0.4$\,s postbounce, $f_{\rm min}$ and $f_{\rm gap}$ have a nearly constant offset of $\sim30$\,Hz.

Note that there is a second local minimum at 600-750\,Hz in the left panel of Fig.~\ref{fig:gap75}, but this does not appear as a power gap in the spectrogram. First, the spectral power is Fourier transformed from the time-domain GW signal with a window of 40\,ms. During this window, the frequency of the second local minimum shifts by $>$10\,Hz, while the change is below 1\,Hz for the observed gap frequency at $\sim$1250\,Hz. Therefore, this second local minimum cannot form a power gap due to interference effects. Second, the spectral power is overall low for frequencies below the peak GW emission, where this second local minimum locates after $\sim0.3$\,s postbounce. So it is harder to be extracted due to the low signal-to-noise ratio.

\begin{figure}
    \centering
    \includegraphics[width=0.48\textwidth]{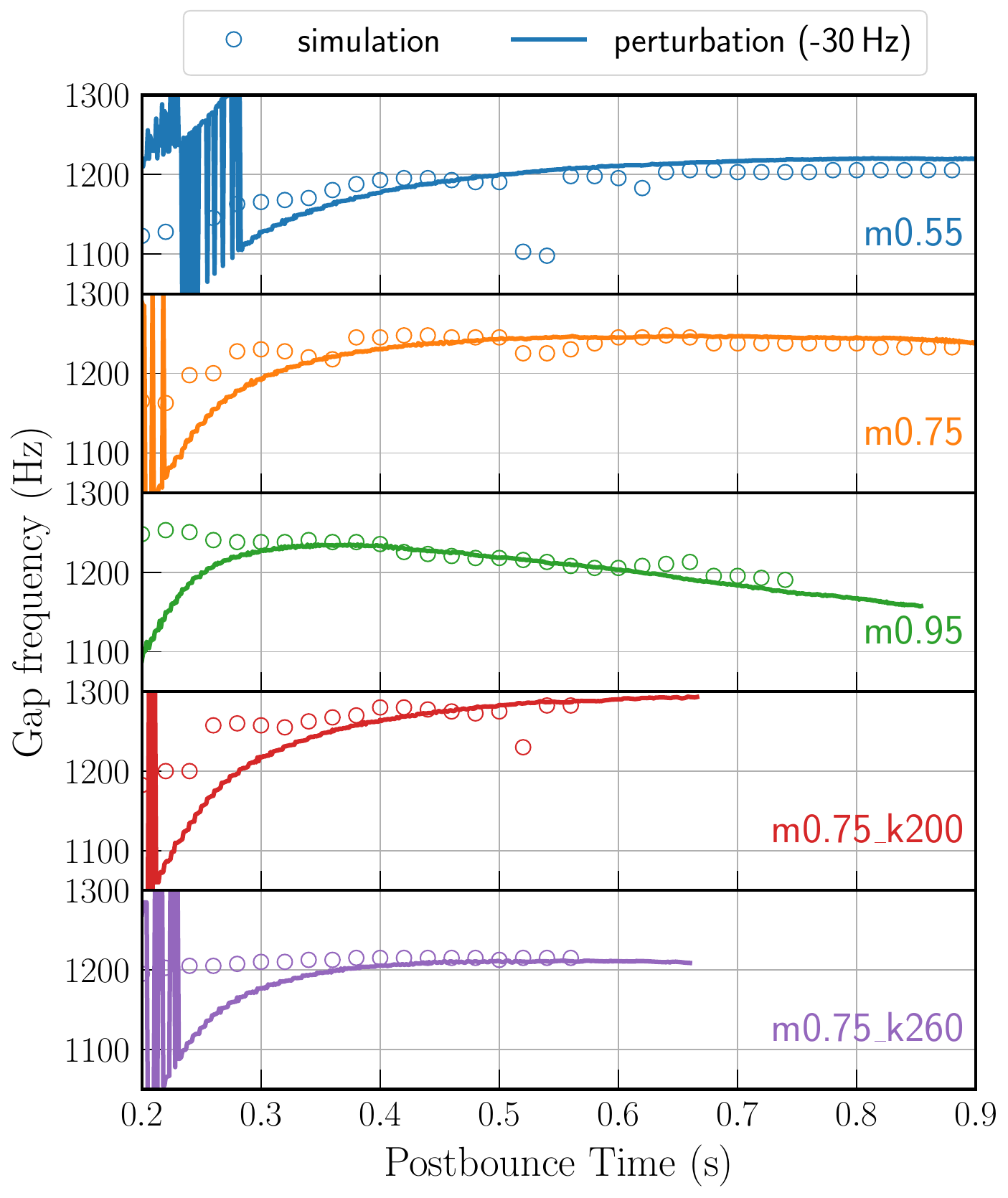}
    \caption{Comparison of gap frequencies in the simulation (open circles) and perturbative analysis (solid lines) for all the supernova models considered in this paper. The gap frequencies from the perturbative analysis are shifted lower by 30\,Hz to better match those from the simulations}
    \label{fig:gap_all}
\end{figure}

We further exemplify the similarity between $f_{\rm min}$ and $f_{\rm gap}$ with other models in Fig.~\ref{fig:gap_all}. The different EOSs result in a different trend for the evolution of $f_{\rm gap}$, most noticeable in the model m0.95 where $f_{\rm gap}$ decreases with time. With $f_{\rm min}$ shifted lower by 30\,Hz, a good agreement is found between $f_{\rm min}$ and $f_{\rm gap}$ for all models after $\sim 0.4$\,s postbounce. We suspect that the 30\,Hz offset results from the numeric differences between the simulation and perturbative analysis. Otherwise, this $f_{\rm min}$ offers a plausible explanation for the power gap of GW emission found in simulations.

We leave a full investigation to future work, but a preliminary investigation into the persistence and the coincident frequency of the power gap between the simulations and the perturbation theory suggests that the term responsible for the overall decrease in the total power surrounding the power gap frequency (between $\sim$1000\,Hz and $\sim$1500\,Hz) in Fig.~\ref{fig:gap75} and therefore the term that enables the zero point in the total power is the $1-N^2/\sigma^2$ term in the differential equation for $\eta_\bot$ (see the $\eta_\bot - \eta_r$ term in Eq.~\ref{eq:matrix}). In particular, the location of its zero point at $N^2 = \sigma^2$ in the inner core, immediately below the convection zone determines the strength of $\eta_\bot$ further out and therefore the overall magnitude of $\delta \hat{\rho}$. Based on the frequency of the zero point in the total power of the perturbative mode functions, Fig.~\ref{fig:gap75}, and preliminary investigations, we suspect that the region where the $N^2 = \sigma_\mathrm{gap}^2$; i.e. $f_\mathrm{BV}=f_\mathrm{min}$, is crucial for this suppression.  This occurs within $\sim$1\,km of the base of the convection zone. 

\section{Conclusions and outlook} \label{sec:conclu}
In this paper, we have applied perturbative analyses for the interpretation of GW signals from CCSNe predicted by 2D pseudo-Newtonian simulations. Our perturbative analyses share the same underlying equations as simulations considered here. By matching the radial profiles of GW emission from the simulation with perturbative functions at \emph{any} frequency and time, we confirmed that global PNS oscillations are responsible for the emission of GWs in CCSNe. Through frequency matching, we tentatively identified the relevant eigenmodes of the PNS oscillation for the peak GW emission to be a set of $g$ modes and the $f$ mode. The switching to the $f$ mode occurs when the peak GW frequency exceeds that of the power gap. Finally, we found that the frequency of the GW power gap within $[1000,1300]$\,Hz coincides with a local minimum in the total GW power of the PNS calculated from perturbative functions. This offers an alternative explanation of the power gap other than the ``trapped" $g$ mode proposed in Ref.~\citep{morozova18}.

Here, we have focused on the emission mechanism of GWs in CCSNe. A further important pending problem is the mechanism of excitation, i.e. what excites the PNS oscillations. \textit{Radice et al.} \citep{radice19} showed the proportional correlation between the energy radiated in GWs and the amount of turbulent energy accreted by the PNS, which suggests the accretion coming from outside excites the PNS oscillations. Other groups \citep{andresen17,mezzacappa20,mezzacappa23} proposed convection inside the PNS as the main driver based on the location of dominant GW emission. The global emission picture drawn from our perturbative analyses may challenge the latter proposal, i.e. the association of location between the emission and excitation of GWs. Nevertheless, we remind the readers that these studies are based on 3D simulations for which one should perform similar analyses as ours to have a conclusive answer.

Despite the challenge of detection due to the rarity of nearby CCSNe, the boost in understanding CCSNe with a potential detection motivates continuous and deeper investigations from the theoretical side. In return, the insights gained from simulations and perturbative analyses can facilitate the development of GW search algorithms that ensure us not missing one of these grand events.

\begin{acknowledgments}
We acknowledge and thank Haakon Andresen for informative discussions.
This work is supported by the National Natural Science Foundation of China (NSFC, Nos. 12288102, 12393811, 12090040/3), the National Key R\&D Program of China (Nos. 2021YFA1600401 and 2021YFA1600400), the International Centre of Supernovae, Yunnan Key Laboratory (No. 202302AN360001). The authors gratefully acknowledge the “PHOENIX Supercomputing Platform” jointly operated by the Binary Population Synthesis Group and the Stellar Astrophysics Group at Yunnan Observatories, CAS. Some computations were enabled by resources provided by the National Academic Infrastructure for Supercomputing in Sweden (NAISS) and the Swedish National Infrastructure for Computing (SNIC) at NSC partially funded by the Swedish Research Council through grant agreements no. 2022-06725 and no. 2018-05973. This work is further supported by the Swedish Research Council (Project No. 2020-00452). 
\end{acknowledgments}

\appendix

\section{Mode function matching at other postbounce times} \label{app:match}
\begin{figure*}
    \centering
    \includegraphics[width=0.95\textwidth]{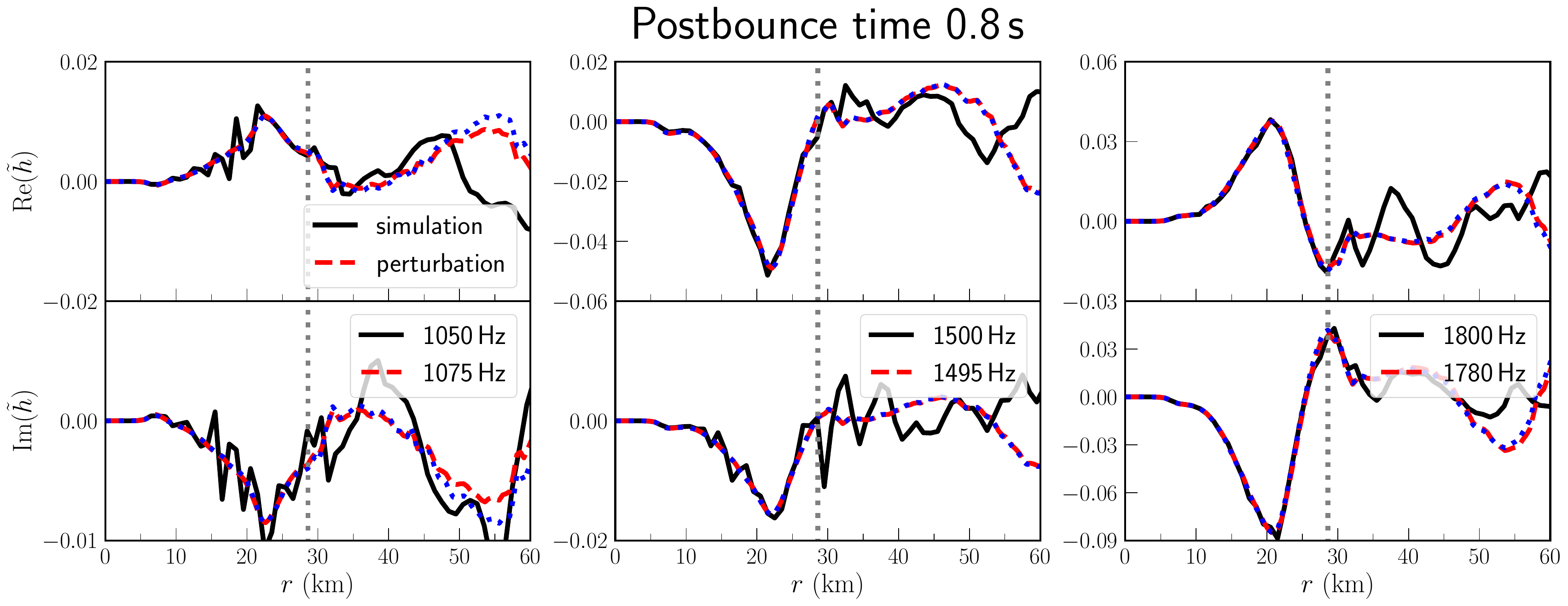} \vspace{0.02\textwidth} \\
    \includegraphics[width=0.95\textwidth]{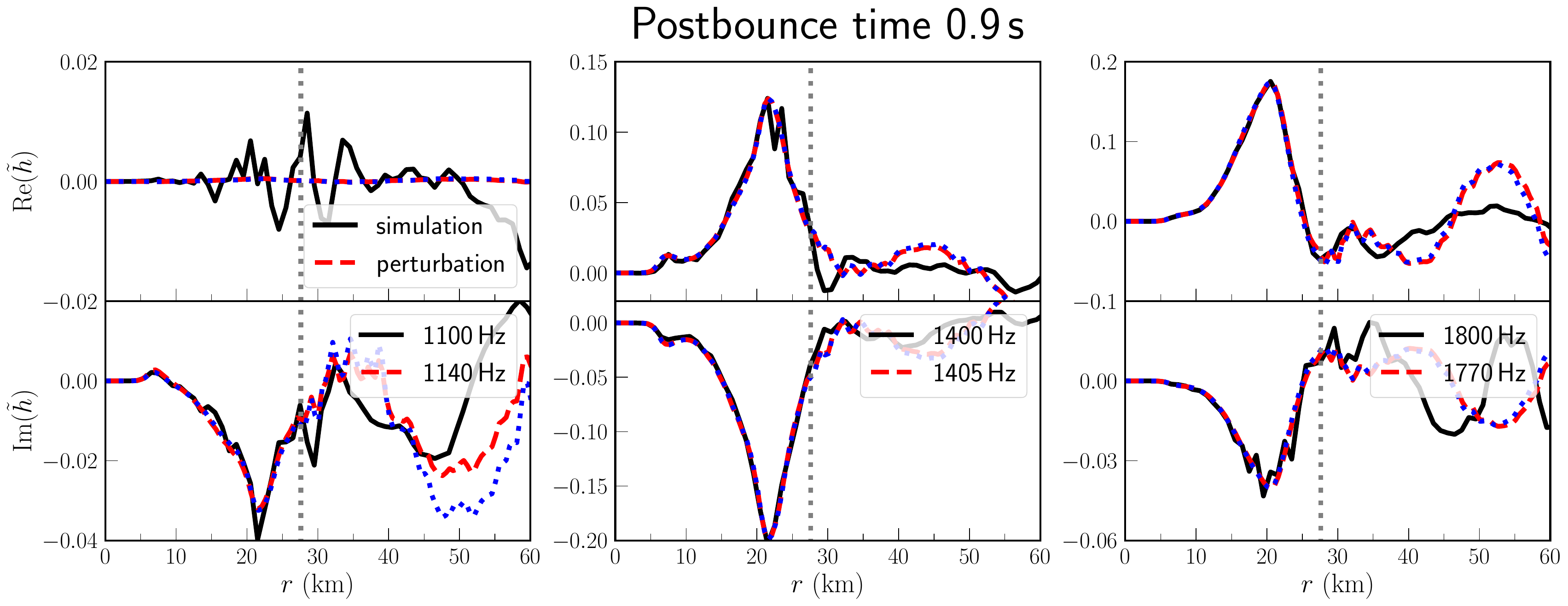}\vspace{0.02\textwidth} \\
    \includegraphics[width=0.95\textwidth]{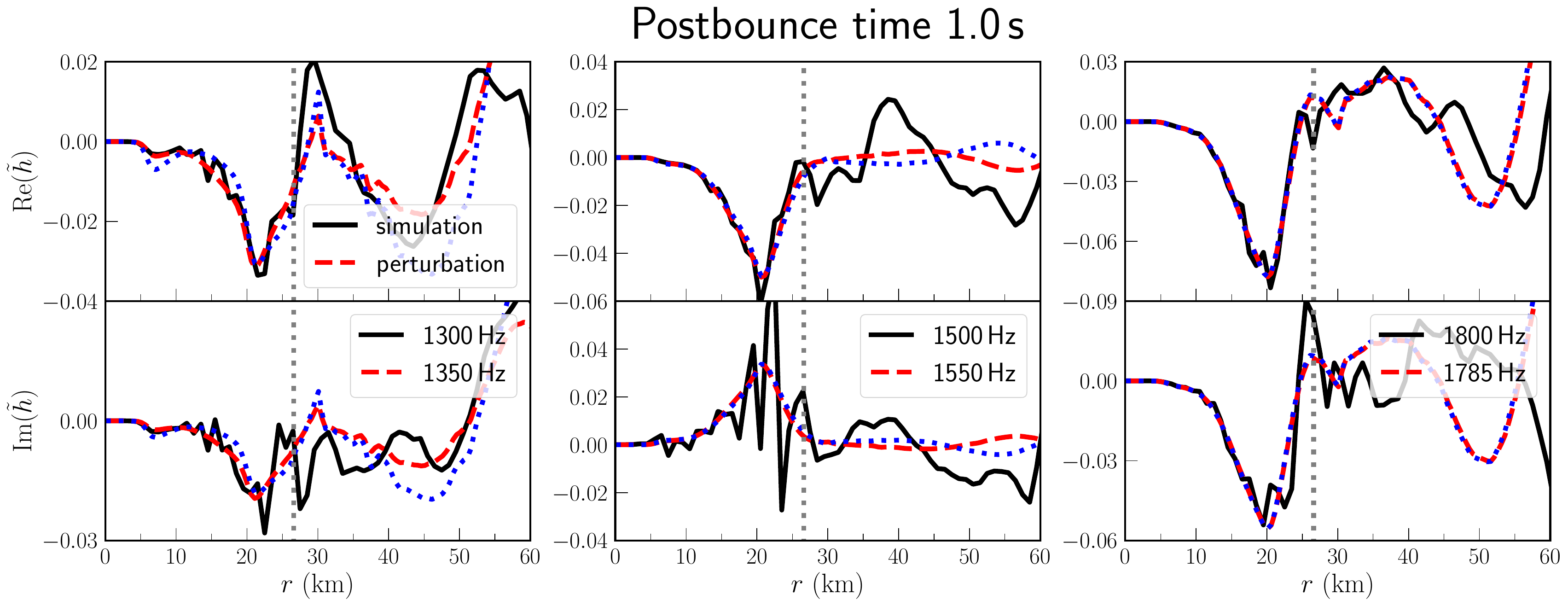}
    \caption{Same as Fig.~\ref{fig:match}, but for the 0.04-s intervals centered at 0.8, 0.9 and 1.0\,s postbounce in the model m0.75 from top to bottom. The blue dotted lines show the perturbative function with the same frequency as the simulation when the best-fit frequency differs from that.}
    \label{fig:app_match}
\end{figure*}

In Fig.~\ref{fig:app_match}, we present the matching of radial profiles of GW emission between the simulation and perturbative mode functions for the latter 3 postbounce times as indicated by the red shaded regions in Fig.~\ref{fig:gw_spec}. Note that the quality of matching becomes worse with smaller GW amplitudes (e.g., the real part of 1100\,Hz at 0.9\,s postbounce), but this does not alter our conclusions.

\section{Frequency matching of eigenmodes in other models} \label{app:eigen}
\begin{figure*}
    \centering
    \includegraphics[width=0.48\textwidth]{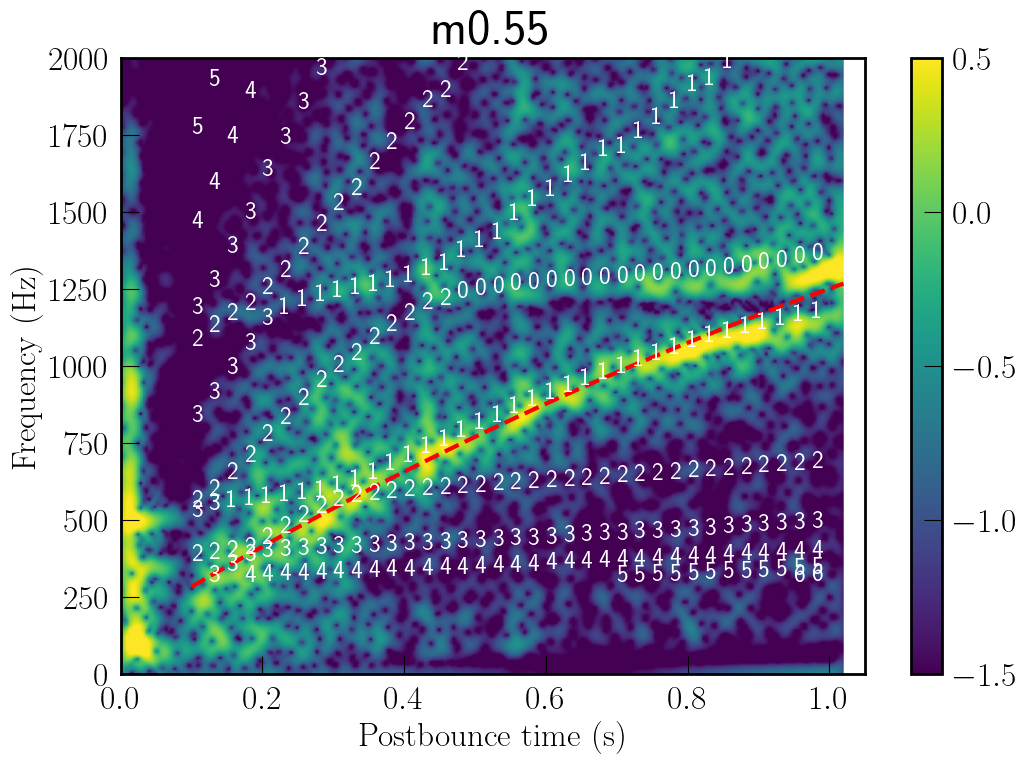}
    \includegraphics[width=0.48\textwidth]{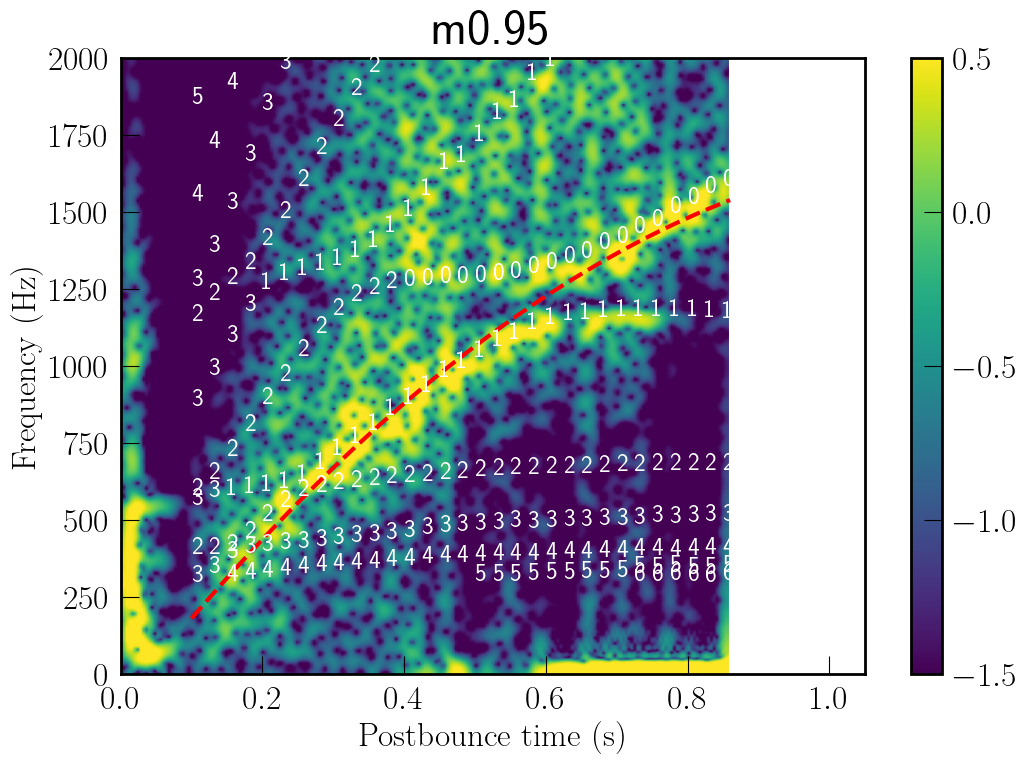}
    \caption{Gravitational-wave spectrograms as Fig.~\ref{fig:gw_spec} but for the models m0.55 (left panel) and m0.95 (right panel). They are over-plotted with a red line denoting the quadratic fitting curve of the peak frequency as a function of time. The frequencies of perturbative eigenmodes are marked with digits that denote the corresponding numbers of radial nodes. \label{fig:spec+other}}
\end{figure*}

In Fig.~\ref{fig:spec+other}, we present the GW spectrograms over-plotted with the frequencies of selected eigenmodes for 2 other models, m0.55 (left panel) and m0.95 (right panel).

\nocite{*}

\bibliography{gw_perturb}

\end{document}